\documentclass[review]{elsarticle}
\usepackage[colorlinks=true,linkcolor=bblue]{hyperref}  

\journal{Applied Mathematical Modelling}

\usepackage{amsmath}	                    
\usepackage{multirow}	                    

\usepackage{color}							
\definecolor{bblue}{rgb}{0.23,0.4,0.7}	    

\renewcommand{\d}{{\rm d}}                  
\newcommand{\jpv}{{j+\frac{1}{2}}}          
\newcommand{\jmv}{{j-\frac{1}{2}}}          
\newcommand{\jp}{{j+1/2}}                   
\newcommand{\jm}{{j-1/2}}                   

\begin{document}

\begin{frontmatter}

\title{A comparison between bottom-discontinuity numerical treatments in the DG framework}

\author[institution1]{Valerio Caleffi\corref{mycorrespondingauthor}}
\ead{valerio.caleffi@unife.it}
\cortext[mycorrespondingauthor]{Corresponding author}
\author[institution1]{Alessandro Valiani}
\ead{alessandro.valiani@unife.it}
\author[institution2]{Gang Li}
\ead{gangli1978@163.com}
\address[institution1]{Department of Engineering, University of Ferrara, Via G. Saragat, 1, 44122 Ferrara, Italy}
\address[institution2]{School of Mathematical Sciences, Qingdao University, Qingdao, Shandong 266071, PR China}

\begin{abstract}
In this work, using an unified framework consisting in third-order accurate discontinuous Galerkin schemes, we perform a comparison between five different numerical approaches to the free-surface shallow flow simulation on bottom steps. 

Together with the study of the overall impact that such techniques have on the numerical models we highlight the role that the treatment of bottom discontinuities plays in the preservation of specific asymptotic conditions. In particular, we consider three widespread approaches that perform well if the motionless steady state has to be preserved and two approaches (one previously conceived by the first two authors and one original) which are also promising for the preservation of a moving-water steady state.

Several one-dimensional test cases are used to verify the third-order accuracy of the models in simulating an unsteady flow, the behavior of the models for a quiescent flow in the cases of both continuous and discontinuous bottom, and the good resolution properties of the schemes. Moreover, specific test cases are introduced to show the behavior of the different approaches when a bottom step interact with both steady and unsteady moving flows.
\end{abstract}


\end{frontmatter}
%
%
\section{Introduction}
Over the last few years, several improvements have been made in the quality of the discontinuous Galerkin (DG) approximations for the nonlinear shallow water equations (SWE). In particular, significant efforts were performed by several researchers to develop numerical techniques for the exact preservation of motionless steady state over non-flat bottom. Because the preservation of the quiescent flow is related to the correct balancing between the flux gradients and the bottom-slope source term the schemes that exactly preserve a stationary flow are denoted as \emph{well-balanced}. The well-balanced property is also referred as \emph{C-property} after the work of Bermudez and Vazquez \cite{BeVa-94}. An updated review on this topic can be found in \cite{XS2014}. For a summary of the well-balancing techniques for two-dimensional DG-SWE schemes the reader is addressed to \cite{Caleffi12} and to the references therein.

Many researchers are now facing further developments of these techniques focusing their attention on the exact preservation of the moving water steady state \cite{Noelle2007,Xing2014wb}. In this case, the exact solution, in absence of discontinuities of the conservative variables (i.e. in absence of bores), is characterized by the constancy of the discharge and of the total head \cite{Murillo2013,Navas2015}. For the latter property of the exact steady solution, a numerical scheme that is able to preserve an initial steady state is defined \emph{energy balanced} in \cite{Murillo2013}.

In parallel to these studies, a relevant effort has been made by many researchers to improve the discretization of balance laws that can not be written in conservative form for the presence of the so-called non-conservative products. These terms make difficult also the simple definition of a correct weak solution if discontinuities are present. A popular theoretical framework (the DLM theory) to deal with such a non-conservative products is due to Dal Maso et al. \cite{DLM}. In this theory, a family of paths linking the states of the conservative variables trough the discontinuity is assumed and properly used to define the weak solutions \cite{DLM}. The DLM theory is successively extended by Par\'{e}s in \cite{PCL} where it is used to construct the \emph{path-conservative} (or path-consistent) family of schemes.

The topics of the well-balancing of numerical schemes and the correct treatment of non-conservative products join when the problem of the consistent modeling of a bottom step in the shallow water framework is faced \cite{Castro2007,Bernetti2008,Caleffi2009,Cozzolino2011}. In fact, the introduction of the trivial equation obtained by equating to zero the time derivative of the bottom elevation allows to write the source term related to the bottom step as a non-conservative product \cite{Gosse2001,Castro2007}. The SWE can be written as an extended system of equations in quasilinear form and the framework proposed by Par\'{e}s \cite{PCL} can be applied to the problem.

An interesting results of this approach is the possibility to introduce a formally correct new definition of well-balanced scheme that allows to take into account the presence of a non-conservative product and the preservation of non-trivial asymptotic steady states. This extended definition can be used when the Jacobian matrix of the system of balance laws has an eigenvalue associated to a linearly degenerate vector field. 
In this context, a numerical method is defined \emph{well-balanced for a given integral curve} related to a linearly degenerate vector field if, given any steady solution belonging to the integral curve, this is preserved at the discrete level \cite{Castro2013}.
For the case we are facing, a numerical model based on the shallow water equations and discontinuous bottom is well-balanced (in the extended sense) if an initial moving steady flow characterized by constant total head and specific discharge is preserved at the discrete level. In particular, we can state that the exact solution over the bottom step is the one that is characterized by constant total energy and specific discharge across the step \cite{LeFloch2007,LeFloch2011,Castro2013}. 

It is interesting also to note that this definition of well-balanced scheme (in the sense of path-conservative schemes) coincides with the the definition of energy balanced scheme in presence of a bottom discontinuity given in \cite{Murillo2013}.

For completeness, it also worth noting that, starting from the observation that the total head throughout a bed discontinuity is not constant in many physical experiments, some researchers propose a different treatments of the bottom step, see for example \cite{Bernetti2008,Cozzolino2011}. A common feature of these different approaches consists in the introduction of semi-empirical expressions for the computation of the resultant of the hydrostatic pressure distribution on the vertical wall of the bottom step. This resultant is successively inserted in a momentum balance related to a control volume that includes the bed discontinuity. Generally these treatment leads to a total head dissipation at the step. 

In this work, we have preferred to follow the idea presented in \cite{LeFloch2007,LeFloch2011,Castro2013,Murillo2013,Navas2015} for its internal consistency with the mathematical properties of the SWE. Therefore we have assumed that the total head has to be constant across the step in steady conditions. This idea is used to obtain the reference solutions for our test cases and to improve certain techniques of literature for the bottom step treatment.

It is worth noting that the numerical approximation of the bottom profile can be discontinuous for two reasons. The bed profile of our test case or application is actually discontinuous, and therefore both the real bottom and its numerical representation are discontinuous, or the bed profile of our test case or application is continuous and only its numerical approximation is discontinuous. In fact, in a DG framework, it is natural that also a continuous bottom is numerically approximated by polynomials that are discontinuous at the cell-interfaces. The techniques for the bottom step management described here are valid for both the cases. Also for this requirement, the idea that the total head has to be preserved in steady conditions across the discontinuity is corrected for our aims.

In this work, a comparison between five different numerical approaches to the flow on bottom steps is performed. The aim is to highlight strengths and weaknesses of the different methods.

First, we consider the simple technique due to Kesserwani and Liang \cite{CKL}. It consists of simplified formulas for the initialization of the bottom data at the discrete level imposing the continuity of the bed profile. While in \cite{CKL} a local linear reconstruction of the bed is suggested, in this work we have tested a parabolic reconstruction to preserve the third-order accuracy. This model is here denoted as the \emph{CKL model}. Then, we take into account the widespread hydrostatic reconstruction method \cite{HSR}, giving rise to the \emph{HSR model}, and a path-conservative scheme \cite{PCL} based on the Dumbser-Osher-Toro (DOT) Riemann solver \cite{DOT} and a linear integration path (giving rise to the \emph{PCL model}).

To treat the discontinuity of the bottom considering a steady moving flow with physically based approaches, the fourth model is obtained modifying the hydrostatic reconstruction scheme as suggested in \cite{Caleffi2009}. This method is characterized by a correction of the numerical flux based on the local conservation of the total head and is here indicated as the \emph{HDR model}.
The fifth, original, model is obtained improving the path-conservative-DOT scheme, through the substitution of the linear integration path with a curved one. The curved path is defined imposing the local conservation of total head, as suggested, albeit in different contexts, in \cite{Pares2004,Castro2007,Mueller13}. The corresponding scheme is the \emph{PCN model}.

The outline of the paper is as follow: in \S~\ref{sec:matmod} the SWE mathematical model is presented in both the conservative and non-conservative form. In \S~\ref{sec:nummod}, after a description of the common elements to all the numerical models, the key elements of each single approach is described. In particular, more space is devoted to the description of the path-conservative model with the non-linear path, provided that the description of the other models can be found in literature. In \S~\ref{sec:res} some test cases are introduced and the different behavior of the models is highlighted. Finally, some conclusions are drawn.
\section{Mathematical Model} \label{sec:matmod}
In this work, the considered balance law consists of the classical nonlinear shallow water equations with the bottom topography source term:
\begin{equation}\label{eq:CSWE}
\partial_t u + \partial_x f = s; \quad \text{with:} \quad u=\begin{bmatrix}h\\ q\end{bmatrix}; \quad
f=\begin{bmatrix}q\\ \frac{q^2}{h} + g\frac{h^2}{2}\end{bmatrix}; \quad
s=\begin{bmatrix}0\\ -g\,h\,\partial_x z\end{bmatrix};
\end{equation}
where: $u(x,t)$, $f(x,t)$ and $s(x,t)$ are the vector of the conservative variables, the flux and the source term, respectively; $h(x,t)$ is the water depth; $q(x,t)$ is the water discharge; $z(x)$ is the bottom elevation; $g$ is the gravity acceleration; $x$ and $t$ are the space and the time, respectively.

To apply the theoretical framework of the path-conservative schemes, Eq.~\eqref{eq:CSWE} is written as a quasi-linear PDE system, introducing the trivial equation $z_t=0$ \cite{Castro2013}:
\begin{equation}\label{eq:PCSWE}
\partial_t w + A(w)\,\partial_x w = 0; \quad \text{with:} \quad w=\begin{bmatrix}h\\ q \\ z \end{bmatrix}; \quad
A(w)=\begin{bmatrix}0&1&0\\c^2-v^2&2v&c^2\\0&0&0\end{bmatrix};
\end{equation}
where: $v=q/h$ is the depth-averaged velocity and $c=\sqrt{g\,h}$ is the relative wave celerity. The matrix $A$ has the the eigenvalues $\lambda_1=v-c$, $\lambda_2=0$ and $\lambda_3=v+c$ and the right eigenvectors, $R_1=[1,\lambda_1,0]^{\text{T}}$, $R_2 = [1, 0, 1-\mathsf{Fr}^2]^{\text{T}}$ and $R_3 = [1,\lambda_2,0]^{\text{T}}$, where $\mathsf{Fr}=|v|/c$ is the Froude number.
\section{The numerical models}\label{sec:nummod}
All the five considered models share common features. First, all the models are integrated in space by a standard discontinuous Galerkin approach using a set of basis for the broken finite element space constituted by \emph{scaled Legendre polynomials} \cite{Xing2014wb}. After the discretization in space, the obtained ODE is integrated using the classical three steps, third-order accurate \emph{strong stability preserving Runge-Kutta} (SSPRK33) scheme \cite{XS2014}. All the integrations on each element and along the paths are performed numerically. To avoid appearance of unphysical oscillations near the solution discontinuities a local limiting procedure is considered \cite{YW_limiter}.
\subsection{The classical conservative models}\label{sec:ccm}
Multiplying the Eq.~\eqref{eq:CSWE} by a polynomial test function $\varphi(x)$, integrating the result over the cell $I_j=[x_\jm,x_\jp]$, applying an integration by parts and the divergence theorem, the \emph{weak formulation} of Eq.~\eqref{eq:CSWE} is obtained:
\begin{equation} \label{eq:weakmatGT}
\int_{I_j} \varphi\, \partial_t u\ \d x - 
\int_{I_j} f\, \partial_x \varphi\ \d x +
\varphi_{\jpv}\, f^{*l}_{\jpv} - \varphi_{\jmv}\, f^{*r}_{\jmv} -
\int_{I_j} \varphi\, s\ \d x
= 0;
\end{equation} 
where $f^{*l}_{\jp}$ and $f^{*r}_{\jm}$ are suitable numerical fluxes, eventually corrected to take into account the bottom discontinuities. A DG numerical approximation of $u(x,t)$ and $z(x)$ in the DG framework is given by:
\begin{equation} \label{eq:uuh}
u^h(x,t) = \sum_{b=1}^{3} \hat{u}_b(t)\: \varphi_b(x) \quad \text{and} \quad z^h(x) = \sum_{b=1}^{3} \hat{z}_b\: \varphi_b(x);
\end{equation} 
where $\{\varphi_b,\ b=1,2,3\}$ is an \emph{orthogonal} basis of the polynomial space of order 3 and $\hat{u}_b(t)$ and $\hat{z}_b$ are the degrees of freedom of the conservative variables and of the bottom elevation, respectively. Here and in the following the superscript $h$ denotes the DG numerical approximations of the variables or the functions evaluated using as arguments the DG numerical approximate variables.

Making the substitution of Eq.~\eqref{eq:uuh} in Eq.~\eqref{eq:weakmatGT} (using a test function $\varphi=\varphi_b$) and introducing the mass matrix $a_b = \int_{I_j} \varphi_b\varphi_b \ \d x$, the following ODE is obtained:
\begin{equation} \label{eq:odeGT}
\frac{\d \hat{u}_b}{\d t} = - \frac{1}{a_b} \left[ 
- \int_{I_j} f^h\, \partial_x \varphi_b\ \d x +
\varphi_{b,\jpv} f^{*l}_{\jpv} - \varphi_{b,\jmv} f^{*r}_{\jmv} -
\int_{I_j} \varphi_b\, s^h\ \d x
 \right];
\end{equation} 
with $b=1,2,3$. Eq.~\eqref{eq:odeGT} represents our numerical models, discretized in space, and it is integrated in time using the SSPRK33 scheme \cite{XS2014}.
\subsubsection{The numerical treatment of the source term and of the flux corrections}\label{sec:fluxcor}
The source term integral in \eqref{eq:odeGT} is computed using standard quadrature starting from the numerical approximation of the source term that ultimately depends on the numerical approximation of the bottom profile \eqref{eq:uuh}. 

Kesserwani and Liang \cite{CKL} proposed simplified formulas for the initialization of the bottom data at the discrete level imposing the continuity of the bed profile at the cell-interfaces. The simplest way to achieve this result consists of assuming as known and unique the bottom elevation at the cell-interfaces, then the bottom is described by the linear segments joining the cell-interface bottom elevations. This approach leads to quite satisfactory results but also clearly reduces the model accuracy to the second order. To avoid this drawback, in the CKL model considered here, we use a parabolic reconstruction of the bottom. We assume as known and unique the values at the cell-interfaces and the cell-averages of the bottom elevation. The three degrees of freedom of the bottom in each cell are computed imposing that the parabolic reconstruction has the prescribed values at the interfaces and the given cell average. In some particular cases, this approach leads to the unphysical lost of monotonicity of the bottom reconstruction. For this reason a check of the reconstructed bottom is performed and, where the monotonicity is lost, the parabolic reconstruction is simply replaced by a straight segment. Because of the continuity of the bottom at the cell-interfaces, the numerical fluxes $f^{*l}_{\jp}$ and $f^{*r}_{\jm}$ are computed without any correction, therefore we have $f^{*l}_{\jp}=f^{*r}_{\jp}=f^{*}_{\jp}$ and $f^{*l}_{\jm}=f^{*r}_{\jm}=f^{*}_{\jm}$.
The HLL approximate Riemann solver $f^*_\jp(u_\jp^-,u_\jp^+)$ is used as numerical flux \cite{XS2014}, where  $u_\jp^-$ and $u_\jp^+$ are computed evaluating at the cell interface $x_\jp$ the approximation $u^h$ relative to the cell $j$ and $j+1$, respectively. The HLL approximate Riemann solver is also used to compute $f^*_\jm$.

Using the scaled Legendre polynomials \cite{Xing2014wb} as basis set, the following equations allow to implement the above described bottom initialization. We focus the attention on the $j$-th cell and we indicate as $\bar{z}_j$, $\tilde{z}_\jp$ and $\tilde{z}_\jm$ the cell-average and the point-values of the bottom, respectively. These quantities are assumed to be known.
The condition that discriminates between monotone and non-monotone solution is: 
\begin{equation}
\left\{\begin{array}{l}
\text{if $x_0 \geq 1$ the solution is monotone} \\
\text{if $x_0 < 1$ the solution is non-monotone} \\
\end{array}   \right.
\end{equation}
where $x_0$ is given be:
\begin{equation}
x_0=\left|\frac{\tilde{z}_\jp - \tilde{z}_\jm}{3\,\left(\tilde{z}_\jp - 2\,\bar{z}_j+ \tilde{z}_\jm\right)} \right|.
\end{equation}
The degrees of freedom of the bottom reconstruction are computed as:
\begin{equation}
\begin{aligned}
\hat{z}_j^1 &= \bar{z}_j;\\
\hat{z}_j^2 &= \tilde{z}_\jp - \tilde{z}_\jm;\\
\hat{z}_j^3 &=
\left\{\begin{array}{ll}
3\,\left(\tilde{z}_\jp - 2\,\bar{z}_j+ \tilde{z}_\jm\right); \quad &\text{if }x_0\geq 1 \\
0; &\text{if } x_0 < 1 \\
\end{array} \right.
\end{aligned}
\end{equation}
The CKL model is well-balanced for the quiescent flow.

The hydrostatic reconstruction \cite{HSR} is a different approach to achieve the well-balancing in the case of motionless steady state, which leads to the HSR model. The degrees of freedom of the bottom are computed through a classical $L_2$ projection and therefore, the bottom profile is piecewise polynomial and discontinuous at the cell interfaces.
The numerical fluxes $f^{*l}_{\jp}$ and $f^{*r}_{\jm}$ are computed as:
\begin{align}
f^{*l}_{\jp} &= f^{*}(u^{*,-}_\jp,u^{*,+}_\jp) + 
\begin{bmatrix} 0\\ \frac{g}{2}\left( h^{-}_\jp\right)^2 - \frac{g}{2}\left( h^{*,-}_\jp\right)^2
\end{bmatrix}; \\
f^{*r}_{\jm} &= f^{*}(u^{*,-}_\jm,u^{*,+}_\jm) + 
\begin{bmatrix} 0\\ \frac{g}{2}\left( h^{+}_\jm\right)^2 - \frac{g}{2}\left( h^{*,+}_\jm\right)^2
\end{bmatrix};
\end{align}
with the left and right values of $u^{*}$ defined by:
\begin{align}
u^{*,\pm}_\jp &= \begin{bmatrix} h^{*,\pm}_\jp \\ h^{*,\pm}_\jp\,v^{\pm}_\jp
\end{bmatrix};\\
h^{*,\pm}_\jp &= h^{\pm}_\jp + z^{\pm}_\jp - \max(z^{+}_\jp,z^{-}_\jp);
\end{align}
where the quantities $h^{\pm}_\jp$, $v^{\pm}_\jp=q^{\pm}_\jp/h^{\pm}_\jp$ and $z^{\pm}_\jp$, representing the approximation of the depth, velocity and bottom elevation, are computed by \eqref{eq:uuh}. The HLL approximate Riemann solver is used as numerical flux for $f^*$. It is interesting to note that the flux correction described here can be interpreted from the physical point of view as the static force exerted by the step on the flow, completely omitting the dynamical effects due to the flow velocity.

The hydrostatic reconstruction approach is used in its original form so in this work further details are not given. The interested reader is addressed to \cite{HSR} for a complete description of the method.

An extension of the hydrostatic reconstruction is proposed in \cite{Caleffi2009}. The approach, that leads to the HDR model, is developed assuming the conservation of the total head on the step in absence of hydraulic jumps and friction terms. First, the total force, $\Phi(u)$, and the specific energy $E(u)$ are introduced as:
\begin{equation}
\Phi(u) = \frac{g\,h^2}{2} + \frac{q^2}{h}; \qquad E(u) = h + \frac{q^2}{2\,g\,h^2}.
\end{equation}
With these functions at hand, the numerical fluxes $f^{*l}_{\jp}$ and $f^{*r}_{\jm}$ are given by:
\begin{align}
f^{*l}_{\jp} &= f^{*}(u^{*,-}_\jp,u^{*,+}_\jp) + 
\begin{bmatrix} 0\\ \Phi\left(u^{-}_\jp\right) - \Phi\left(u^{*,-}_\jp\right)
\end{bmatrix};  \\
f^{*r}_{\jm} &= f^{*}(u^{*,-}_\jm,u^{*,+}_\jm) + 
\begin{bmatrix} 0\\ \Phi\left(u^{+}_\jm\right) - \Phi\left(u^{*,+}_\jm\right)
\end{bmatrix};
\end{align}
where the quantities $u^{*,-}_\jp$ and $u^{*,+}_\jm$ are computed as follows. Without loss of generality the attention is focused on $u^{*,-}_\jp$.
We introduce a virtual section between the $j$-th and the $j+1$-th cells and a virtual layer of infinitely small length between the interface at $x_\jp$ of the $j$-th cell and the virtual section. The bottom elevation for the virtual section is $z_\jp^*=\max(z_\jp^-,z_\jp^+)$. Then we compute:
\begin{equation}
E_\jp^{*,-}=(z_\jp^{-} - z_\jp^{*}) + E_\jp^{-};
\end{equation}
with $E_\jp^{-} = E(u_\jp^{-})$. This relation is obtained imposing the conservation of the total head and the of the discharge into the virtual layer. $E_\jp^{*,-}$ can be interpreted as the function $E$ computed at the virtual section at $x_\jp^-$ location, i.e.:
\begin{equation}\label{eq:espi}
E_\jp^{*,-} = h_\jp^{*,-} + \frac{(q_\jp^{-})^2}{2\,g(h_\jp^{*,-})^2}
\end{equation}
that corresponds to an implicit expression for $h_\jp^{*,-}$. 
Finding the approximate value of $h_\jp^{*,-}$ that satisfies Eq.~\eqref{eq:espi} for given values of $q_\jp^{-}$ and $E_\jp^{*,-}$ using numerical techniques is straightforward. Notwithstanding this, a better choice is to solve the problem analytically using the solution proposed in \cite{Valiani2008}. In fact, such solution is exact, and not approximate, and also much more efficient because any iteration is avoided. Two depths make physical sense and correspond to a subcritical and a supercritical solution. In this work we select the solution consistent with the Froude number $\mathsf{Fr}(u_\jp^-)$: 
\begin{equation}
h_\jp^{*,-}=\left\{\begin{array}{lll}
\left.\left(E_\jp^{*,-}\right)^{-1}\right|_{\text{subcritical}} & \text{if} & \mathsf{Fr}(u_\jp^-)<1;\\[4ex]
\left.\left(E_\jp^{*,-}\right)^{-1}\right|_{\text{supercritical}} & \text{if} & \mathsf{Fr}(u_\jp^-)>1;
\end{array}\right.
\end{equation}

Once calculated $h_\jp^{*,-}$ we can write $u^{*,-}_\jp=[h_\jp^{*,-},q_\jp^{-}]^\text{T}$. 

The flux correction can be interpreted from the physical point of view as the resultant of both the static and the dynamic forces exerted by the step on the flow. The reader is addressed to \cite{Caleffi2009} for further details.

%
\subsection{The path-conservative models}\label{sec:pcm}
To construct a path-conservative DG scheme, the weak formulation of the quasi-linear system \eqref{eq:PCSWE} is obtained multiplying Eq.~\eqref{eq:PCSWE} by a polynomial test function $\varphi(x)$ and integrating the result over the cell $I_j$:
\begin{equation} \label{eq:weakmat}
\int_{I_j} \varphi\, \partial_t w\ \d x + 
\int_{I_j} \varphi\, A\, \partial_x w\ \d x +
\varphi_{\jpv} \mathcal{D}^{-}_{\jpv} + \varphi_{\jmv} \mathcal{D}^{+}_{\jmv}
= 0;
\end{equation} 
where $\mathcal{D}^{\pm}_{\jp}$ are the fluctuations between the cells \cite{PCL}.
The DG approximation of $w(x,t)$ is introduced as:
\begin{equation} \label{eq:uh}
w^h(x,t) = \sum_{b=1}^{3} \hat{w}_b(t)\: \varphi_b(x);
\end{equation} 
where $\{\varphi_b,\ b=1,2,3\}$ is the basis of the polynomial space of order 3 and $\hat{w}_b$ are the degrees of freedom of the dependent variables.
Substituting Eq.~\eqref{eq:uh} in Eq.~\eqref{eq:weakmat} (using a test function $\varphi=\varphi_b$) and introducing the mass matrix $a_b = \int_{I_j} \varphi_b\varphi_b \ \d x$, with some algebra, we obtain:
\begin{equation} \label{eq:ode}
\frac{\d \hat{w}_b}{\d t} = - \frac{1}{a_b} \left[ 
\int_{I_j} \varphi_b\, A^h\, \partial_x w^h\ \d x +
\varphi_{b,\jpv} \mathcal{D}^{-}_{\jpv} + \varphi_{b,\jmv} \mathcal{D}^{+}_{\jmv} \right];
\end{equation} 
with $b=1,2,3$. The fluctuations $\mathcal{D}^{\pm}_{\jp}$, depending from the discontinuous values $w_{\jp}^{-}$ and $w_{\jp}^{+}$ at the cell interfaces $x_\jp$, are computed using the Dumbser-Osher-Toro (DOT) Riemann solver \cite{DOT}:
\begin{equation} \label{eq:Dpm}
\mathcal{D}^{\pm}_{\jpv} = \frac{1}{2}\int_0^1 \left[
A(\Psi(w_{\jpv}^{-},w_{\jpv}^{+},s)) \pm 
\left|A(\Psi(w_{\jpv}^{-},w_{\jpv}^{+},s)) \right|\right]\frac{\partial \Psi}{\partial s}\ \d s;
\end{equation} 
where the absolute-value matrix-operator is defined by $|A| = R\,|\Lambda|\, R^{-1}$ with $|\Lambda|=\text{diag}(|\lambda_1|,|\lambda_2|,|\lambda_3|)$ and $R$ the right eigenvectors matrix. The choice of the integration path $\Psi(w_{\jp}^{-},w_{\jp}^{+},s)$, given as a parametrized function of $s \in [0,1]$ is of fundamental importance in the behavior of the model.

Eqs.~\eqref{eq:ode}-\eqref{eq:Dpm} represent the semi-discretized form of the path-conservative models and it is integrated in time using the SSPRK33 scheme \cite{XS2014}.

\subsubsection{The choice of the integration path}\label{sec:path}
Indicating with $w^{-}$ and $w^{+}$ the values of the variables before and after a cell interface the fluctuations $\mathcal{D}^{\pm}$ for the given path $\Psi(s)$ have to satisfy the two relations:
\begin{align}
\mathcal{D}^{-}\left(w^-,w^+,\Psi(s)\right) + \mathcal{D}^{+}\left(w^-,w^+,\Psi(s)\right) 
&= \int_0^1 A(\Psi(w^-,w^+,s)) \frac{\partial \Psi}{\partial s} \d s, \\
\mathcal{D}^{\pm}\left(w,w,\Psi(s)\right) &= 0;
\end{align}
where the path $\Psi=\Psi(w^-,w^+,s)$ is a continuous function in the parameter $s \in [0,1]$ that connects the states $w^-$ and $w^+$ in the phase space, satisfying $\Psi(w^-,w^+,0) = w^-$ and $\Psi(w^-,w^+,1) = w^+$ \cite{PCL}.

Working on the SWE, the use of a simple linear path, $\Psi(w^-,w^+,s)= w^- + s(w^+ - w^-)$, is sufficient to obtain reasonable results only if the motionless steady state has to be preserved. The use of this path in Eq.~\eqref{eq:Dpm} leads to the PCL model.

In order to improve the treatment of the moving water steady states, a different path, 
inspired by previous works \cite{Pares2004,Castro2007,Mueller13}, is here introduced.

First, the attention is focused on the integral curve in the phase space, $\gamma_{LD}(s)$, of the linearly degenerate (LD) vector field $R_2$ associated to the eigenvalue $\lambda_2=0$. 
The corresponding \emph{generalized Riemann invariants} $\Gamma^{LD}$ (i.e., the functions of $w$ whose values are invariant along $\gamma_{LD}(s)$) are:
\begin{equation}
\Gamma^{LD}_1 = q = \tilde{q}; \qquad
\Gamma^{LD}_2 = H = z+E = z+h+\frac{q^2}{2\,g\,h^2} = \tilde{H};\label{eq:param2}
\end{equation}
where $\tilde{q}$ and $\tilde{H}$ are two real constants \cite{Castro2013}. It is worth noting that $H$ is the total head and therefore, on the basis of Eq.~\eqref{eq:param2} we can state that \emph{the total head and the specific discharge are constant along the integral curve}.

In the context of the path-conservative schemes, a numerical method is defined \emph{well-balanced} for $\gamma_{LD}$ if, given any moving-water steady solution $w^{(s)}(x) \in \gamma_{LD}, \forall x \in (x_l,x_r) \subset \mathcal{R}$ and an initial condition $w^h_j \in \gamma_{LD}, \forall j \in [1,\ldots,N]$, where $N$ is the number of cells used to discretize the domain, than the initial state is preserved \cite{Castro2013}. This general definition implies that to construct a well-balanced model, all the three terms on the RHS of Eq.~\eqref{eq:ode} (the integral and the fluctuations) have to be zero if they are computed for a steady solution characterized by constant total head and discharge. In this work we focus our attention only on the fluctuations, $\mathcal{D}^{\pm}$, ignoring the integral of Eq.~\eqref{eq:ode}, achieving the well-balancing only in particular conditions (e.g. piecewise constant solutions).

If we use a path $\Psi(w^-,w^+,s)$ that, in steady conditions, with $w^\pm \in \gamma_{LD}$, is a parametrization of the arc of $\gamma_{LD}$ the following obvious relation holds: 
\begin{equation}\label{eq:pathcurv}
\frac{\partial \Psi}{\partial s} = \gamma’_{LD}.
\end{equation}
Moreover, by definition, at each point $\gamma_{LD}(s)$ the tangent vector $\gamma’_{LD}(s)$ is an eigenvector of $A(\gamma_{LD}(s))$  corresponding to the eigenvalue $\lambda_2(\gamma_{LD}(s))$, i.e.:
\begin{equation}\label{eq:Adpsids}
A(\gamma_{LD}(s))\; \gamma’_{LD}(s) = \lambda_2(\gamma_{LD}(s))\; \gamma’_{LD}(s) = 0.
\end{equation}
Substituting Eq.~\eqref{eq:pathcurv} in Eq.~\eqref{eq:Adpsids} and using basic algebraic manipulation, we have:
\begin{equation}
A(\gamma_{LD}(s)) \left.\frac{\partial \Psi}{\partial s}\right|_{\gamma_{LD}(s)} = 0;\qquad
|A(\gamma_{LD}(s))| \left.\frac{\partial \Psi}{\partial s}\right|_{\gamma_{LD}(s)} = 0;\label{eq:Adpds}
\end{equation}
and it is easy to check that Eqs.~\eqref{eq:Adpds} lead to $\mathcal{D}^{\pm}=0$. In other words if a path that in steady conditions corresponds to an arc of $\gamma_{LD}$, the corresponding numerical scheme \eqref{eq:ode}-\eqref{eq:Dpm} is well-balanced in the sense of the path-conservative schemes.

A path $\Psi(w^-,w^+,s)$ satisfying such condition in a steady state is:
\begin{equation}
\Psi(s) = \begin{bmatrix}
\bar{h}(s)   \\ 
\bar{q}(s)   \\
\bar{z}(s)   \end{bmatrix}=
\begin{bmatrix}
\bar{E}(s)^{-1}  \\ 
q^-_\jp + s (q^+_\jp - q^-_\jp)   \\
z^-_\jp + s (z^+_\jp - z^-_\jp)   \end{bmatrix};
\end{equation}
with:
\begin{equation}\label{eq:Espec}
\bar{E}(s) = \bar{h}(s)+\frac{[\bar{q}(s)]^2}{2\,g\,[\bar{h}(s)]^2} = \bar{H}(s) - \bar{z}(s);
\end{equation}
and:
\begin{equation}
\bar{H}(s) = \bar{H}^-_{\jp} + s (\bar{H}^+_{\jp} - \bar{H}^-_{\jp}).
\end{equation}

The computation of $\bar{E}^{-1}$, i.e.\ finding the values of $\bar{h}$ that satisfy \eqref{eq:Espec}, is again performed analytically \cite{Valiani2008}. We have considered only the cases where both $w^-$ and $w^+$ are subcritical or supercritical. The other cases require further investigations that are beyond the scope of this work. 
\subsection{Relationship between flux corrections and integration paths}
To obtain the desired properties working with the conservative models \S~\ref{sec:ccm}, the attention is focused on the flux correction that takes into account the action exerted on the flow by the step. On the other hand, to reach the same aim in the context of the path-conservative models of \S~\ref{sec:pcm}, we have focused our attention on the path selection. Notwithstanding the differences between the two approaches a strong relationship between them exists. The following reasoning is useful to highlight the similarities in the two approaches.

It is worth noting that the behavior of the flow at the bottom step is related to the wave associated to the eigenvalue $\lambda_2$ of the matrix $A$ given in Eq.~\eqref{eq:PCSWE}. This wave is a contact discontinuity related to a linearly degenerate vector field \cite{Pares2004,Castro2007}. A relationship between the flow state before the bottom step, $w^-$, and after the bottom step, $w^+$, can be written requiring that the \emph{generalized Rankine Hugoniot condition} has to be satisfied \cite{LeFloch2007,LeFloch2011,Berthon2015}:
\begin{equation}\label{eq:grh}
\xi\left(w^+ - w^-\right) = \int_0^1 A(\Psi(w^-,w^+,s)) \frac{\partial \Psi}{\partial s} \d s;
\end{equation}
where $\xi$ is the contact wave celerity, $A$ is the matrix given in Eq.~\eqref{eq:PCSWE} and $\Psi$ is the selected path. Provided that $\lambda_2$ is always equal to zero and therefore the contact wave is stationary (i.e. $\xi = 0$), Eq.~\eqref{eq:grh} becomes:
\begin{equation}\label{eq:grh0}
\int_0^1 A(\Psi(w^-,w^+,s)) \frac{\partial \Psi}{\partial s} \d s = 0;
\end{equation}

The matrix $A(w)$ can now be split in two parts:
\begin{equation}\label{eq:asplit}
A(w) = A_c(w) + A_n(w);
\end{equation}
with:
\begin{equation}
A_c(w)=\begin{bmatrix}0&1&0\\c^2-v^2&2v&0\\0&0&0\end{bmatrix}
\quad \text{and:}\quad A_n(w)=\begin{bmatrix}0&0&0\\0&0&g\,h\\0&0&0\end{bmatrix};
\end{equation}
where the matrix $A_c$ is clearly the Jacobian of a the flux $\hat{f}=\left[h\,v, h\,v^2 + g\,h^2/2, 0\right]^\text{T}$. In other words, we can formally write $A_c=\partial \hat{f} / \partial w$.

The substitution of \eqref{eq:asplit} into the Eq.~\eqref{eq:grh0} leads to:
\begin{equation}\label{eq:grh0s}
\int_0^1 A_c(\Psi(w^-,w^+,s)) \frac{\partial \Psi}{\partial s} \d s +
\int_0^1 A_n(\Psi(w^-,w^+,s)) \frac{\partial \Psi}{\partial s} \d s = 0;
\end{equation}

The first integral of \eqref{eq:grh0s} can be analytically computed and the result is independent from the choice of the path. We obtain:
\begin{equation}\label{eq:grh0sf}
\hat{f}^+ - \hat{f}^- + \int_0^1 A_n(\Psi(w^-,w^+,s)) \frac{\partial \Psi}{\partial s} \d s =0;
\end{equation}
where $\hat{f}^-$ and $\hat{f}^-$ are the fluxes before and after the step, respectively.

The second integral of \eqref{eq:grh0s} affects only the second component of \eqref{eq:grh0s} that can be written in an explicit form as:
\begin{equation}\label{eq:grhexp}
[hv^2+gh^2/2]^+ - [hv^2+gh^2/2]^- + \int_0^1 g\,h|_\Psi \frac{\partial z}{\partial s} \d s = 0.
\end{equation}

From the physical point of view Eq.~\eqref{eq:grhexp} is a momentum balance where the latter term represents the force exerted by the step on the flow. The latter term is the only one depending from the path. 
We can conclude that the choice of the integration path corresponds to the choice of an estimate of the forces exerted by the step on the flow.

\section{Results}\label{sec:res}
Several test cases are used to validate each aspect of the models and to perform the comparison between them. Only the more interesting results are here reported.

\subsection{C-property test case}
The purpose of this test case is to verify the fulfillment of the C-property over a non-flat bottom \cite{BeVa-94}. To verify the C-property on smooth and discontinuous bottoms using only one test case we have introduced an original bed profile. This profile is generally continuous (i.e.\ differentiable) but with two discontinuities. Moreover, to make the test reliable for any bathymetry, the analytical function describing the bed profile is defined using harmonic functions instead of polynomial functions in order to avoid the unintended exact correspondence between the test case bottom and the corresponding numerical approximation. 

\begin{figure}[h]
\begin{center}
\includegraphics[width=12cm]{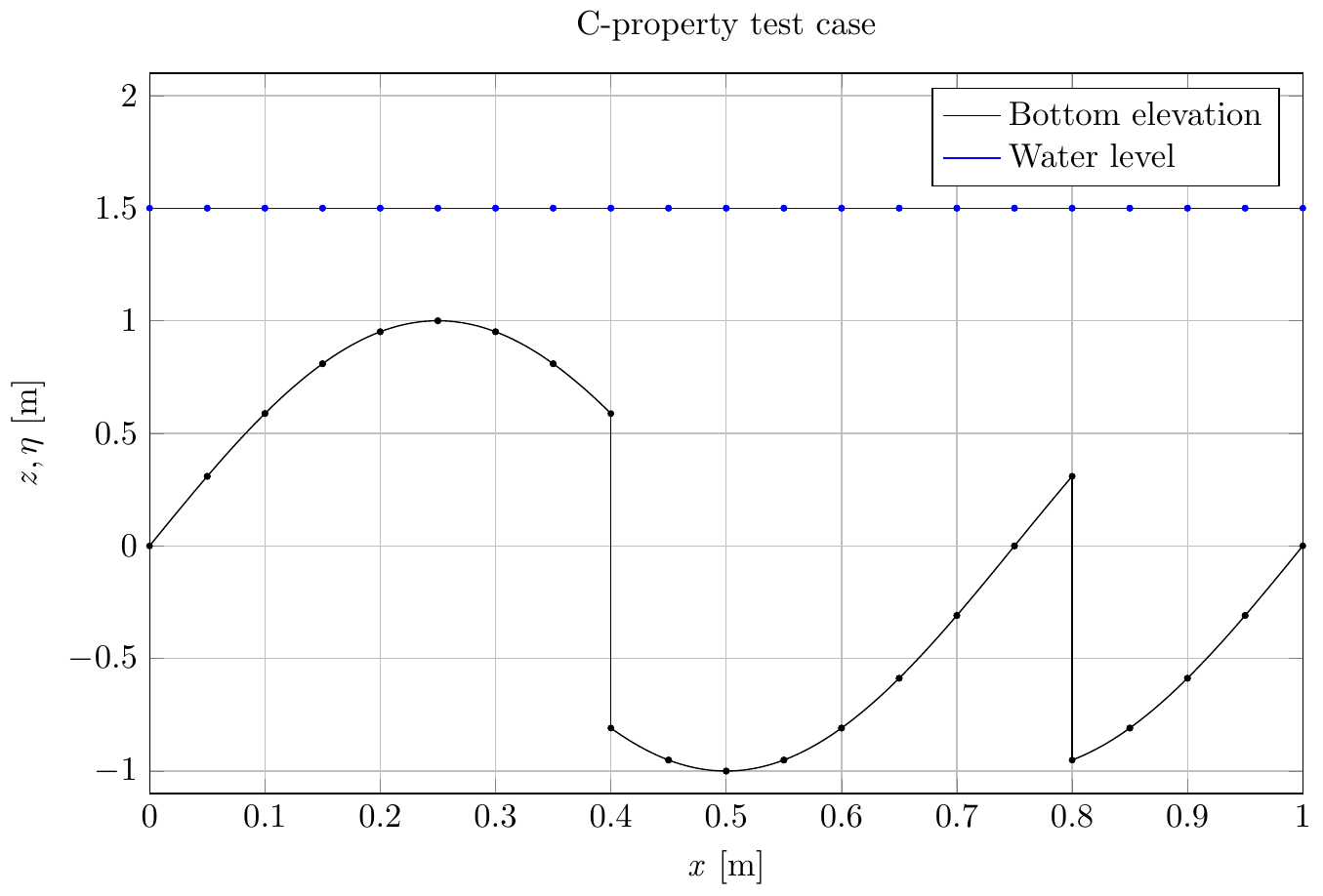}
\end{center}
\caption{C-property test case: water level and bottom profile.}\label{fig:cp}
\end{figure}

The bottom profile is:
\begin{equation}
z(x)=\left\{\begin{array}{lll}
\sin(2\pi x) & \text{if} & 0.0 \text{ m} < x \leq 0.4 \text{ m};\\
\cos(2\pi (x+1)) & \text{if} & 0.4 \text{ m} < x < 0.8 \text{ m};\\
\sin(2\pi x) & \text{if} & 0.8 \text{ m} < x \leq 1.0 \text{ m};
\end{array}\right.
\end{equation}
A constant free-surface elevation, $\eta$ = 1.5 m, and a zero discharge are the initial conditions. The boundary conditions are periodic. To test the ability of the schemes to maintain the initial quiescent flow, simulations are carried out until $t$ = 0.1 s, using a mesh of 20 cells. Fig.~\ref{fig:cp} shows the bottom profile and the water level.

The $L_1$, $L_2$ and $L_{\infty}$ norms of the errors of the water level and the specific discharge are computed. The results, obtained using the double precision floating-point arithmetics in numerical computations, are summarized in Tab.~\ref{tab:C-property}. 
\begin{table} 
\centering
\begin{small}
\begin{tabular}{c c c c | c c c} 
\hline 
\hline 
       & \multicolumn{3}{c|}{$\eta$} & \multicolumn{3}{c}{$q$}\\
\hline 
Model & $L_1$ & $L_2$ & $L_{\infty}$ & $L_1$ & $L_2$ & $L_{\infty}$\\ 
\hline \hline 
CKL & 3.4417e-16 & 4.4686e-16 & 8.8818e-16 & 2.9854e-15 &  3.8310e-15 & 8.9294e-15 \\
HSR & 4.4409e-16 & 5.6610e-16 & 1.1102e-15 & 3.0346e-15 &  4.0246e-15 & 1.2050e-14 \\
PCL & 3.3307e-16 & 4.3850e-16 & 8.8818e-16 & 7.0776e-15 &  1.1399e-14 & 2.9616e-14 \\
PCN & 2.5535e-16 & 3.6822e-16 & 8.8818e-16 & 7.2530e-15 &  1.0642e-14 & 3.3617e-14 \\
HDR & 3.8858e-16 & 5.9374e-16 & 1.9984e-15 & 3.9827e-15 &  5.5392e-15 & 1.3453e-14 \\
\hline 
\hline 
\end{tabular}
\end{small}
\caption{C-property test case: the $L_1$, $L_2$ and $L_{\infty}$ norms of the errors in terms of water elevation $\eta$ and specific discharge $q$, for a motionless steady flow, are shown for all the five models.} \label{tab:C-property} 
\end{table} 
The differences of the numerical solutions from the reference solution are only due to round-off errors. These results prove the fulfillment of the exact C-property for all the models considered in this work.

\subsection{Accuracy Analysis}
The space and time accuracy of the scheme is verified using the test case proposed by Xing and Shu \cite{Xing-05} concerning a smooth unsteady flow. The bottom is given by $z(x)=\sin^2(\pi x)$, while the initial conditions are:
\begin{equation}
h(x,0) = h_0 + \text{e}^{\cos(2 \pi x)};\qquad
q(x,0) = \sin(\cos(2 \pi x));
\end{equation}
with $h_0$=5 m and $x \in [0,1]$ m. Periodic boundary conditions are assumed and the duration of the simulation is $t$ = 0.1 s. The accuracy analysis is performed using as a reference the numerical solution computed on a very fine mesh of 6561 cells. In Tab.~\ref{tab:accuracy_Level} the $L_1$, $L_2$ and $L_{\infty}$ norms of the errors and the corresponding order of accuracy, for the water level, are reported. The third-order accuracy is achieved for any norm and for any model confirming that the accuracy of the schemes agrees with the expected one. It is interesting to note that the use of a sub-optimal linear reconstruction of the bottom profile in the CKL model gives rise to a loss of accuracy. In this case, the CKL model is only second-order accurate. 
\begin{table}
\centering
\begin{small}
\begin{tabular}{c r c c c c c c}
  \hline 
  \hline 
  Model & Cells & $L_1$ & order & $L_2$ & order & $L_{\infty}$ & order\\ 
  \hline
\multirow{4}{*}{CKL}
   &     81 & 2.0249e-05 &         & 5.2464e-05 &         &  3.7674e-04 &         \\ 
{} &    243 & 6.5155e-07 &  3.1280 & 1.6297e-06 &  3.1601 &  1.0369e-05 &  3.2702 \\ 
{} &    729 & 2.4288e-08 &  2.9941 & 7.2609e-08 &  2.8318 &  6.8390e-07 &  2.4747 \\ 
{} &   2187 & 8.8408e-10 &  3.0158 & 3.4708e-09 &  2.7678 &  6.1255e-08 &  2.1962 \\ 
\hline
\multirow{4}{*}{HSR}
   & 	 81 & 1.5957e-05 &         & 5.0270e-05 &         &  3.7801e-04 &         \\ 
{} &    243 & 4.9341e-07 &  3.1643 & 1.3600e-06 &  3.2859 &  1.0433e-05 &  3.2677 \\ 
{} &    729 & 1.8288e-08 &  2.9993 & 4.9587e-08 &  3.0143 &  3.7883e-07 &  3.0180 \\ 
{} &   2187 & 6.6220e-10 &  3.0206 & 1.7689e-09 &  3.0342 &  1.3455e-08 &  3.0381 \\ 
\hline
\multirow{4}{*}{PCL}
   &     81 & 1.5174e-05 &         & 4.8063e-05 &         &  3.5963e-04 &         \\ 
{} &    243 & 4.9720e-07 &  3.1115 & 1.3582e-06 &  3.2462 &  1.0486e-05 &  3.2177 \\ 
{} &    729 & 1.8540e-08 &  2.9939 & 5.0337e-08 &  2.9994 &  3.8364e-07 &  3.0112 \\ 
{} &   2187 & 6.8050e-10 &  3.0082 & 1.8007e-09 &  3.0316 &  1.3720e-08 &  3.0319 \\ 
\hline
\multirow{4}{*}{PCN}
   &     81 & 1.5174e-05 &         & 4.8063e-05 &         &  3.5963e-04 &         \\ 
{} &    243 & 4.9720e-07 &  3.1115 & 1.3582e-06 &  3.2462 &  1.0486e-05 &  3.2177 \\ 
{} &    729 & 1.8540e-08 &  2.9939 & 5.0337e-08 &  2.9994 &  3.8364e-07 &  3.0112 \\ 
{} &   2187 & 6.8050e-10 &  3.0082 & 1.8007e-09 &  3.0316 &  1.3720e-08 &  3.0319 \\ 
\hline 
\multirow{4}{*}{HDR}
   &     81 & 1.5955e-05 &         & 5.0269e-05 &         &  3.7801e-04 &         \\ 
{} &    243 & 4.9332e-07 &  3.1643 & 1.3600e-06 &  3.2859 &  1.0433e-05 &  3.2677 \\ 
{} &    729 & 1.8285e-08 &  2.9993 & 4.9587e-08 &  3.0143 &  3.7883e-07 &  3.0180 \\ 
{} &   2187 & 6.6206e-10 &  3.0206 & 1.7689e-09 &  3.0342 &  1.3455e-08 &  3.0382 \\ 
  \hline
  \hline 
\end{tabular}
\end{small}
\caption{Accuracy analysis: the $L_1$, $L_2$ and $L_{\infty}$ norms of the errors and the corresponding accuracy orders in terms of water elevation are shown for all the models.}
\label{tab:accuracy_Level}
\end{table}
\subsection{Riemann problem with a bottom step}
\begin{figure}
\centering
\includegraphics[width=10cm]{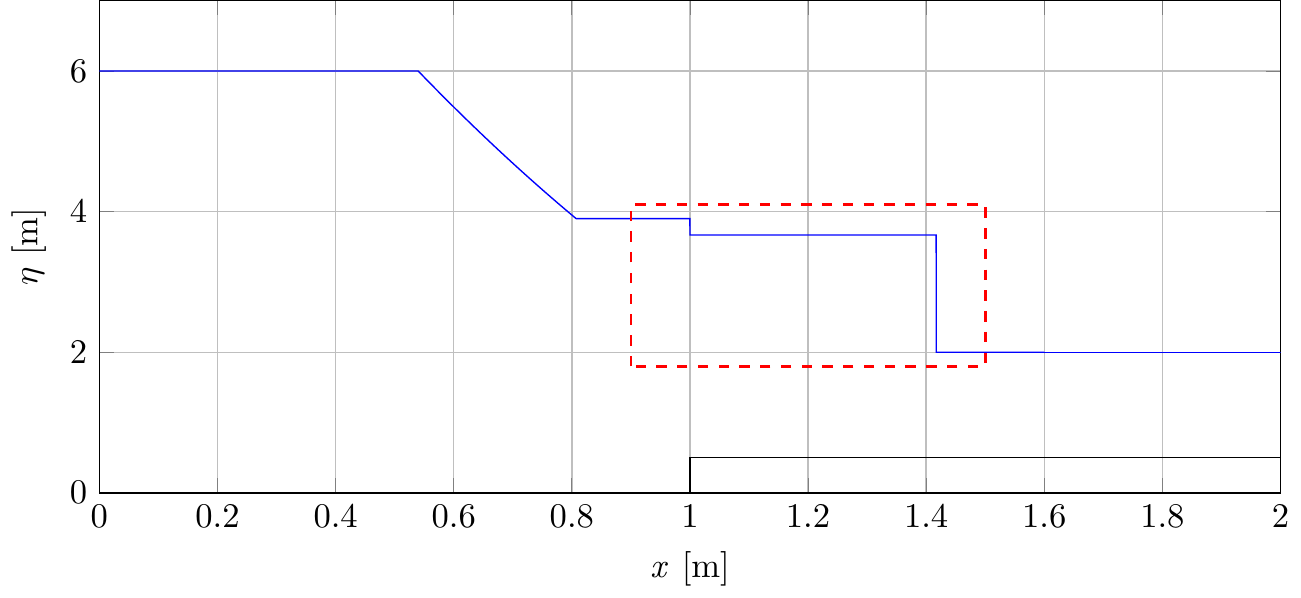}
\caption{Riemann problem with a bottom step: the solution consists of a rarefaction, a stationary contact wave and a shock. Only the results inside the red rectangle are shown in Fig.~\ref{fig:SS_zoom}.}
\label{fig:SS_soluzione}
\end{figure}

This test case, constituted by an initial values problem with piecewise initial data, is used to verify the behavior of numerical models in the reproduction of an unsteady flow. In particular, the shock-capturing properties of the five models are highlighted by the presence of a moving discontinuity in the reference solution. 
The channel is 2 m long and the bottom elevation is zero for $x<1$ m and 0.5 m for $x>1$ m. The initial free-surface level is 6 for m $x<1$ m and 2 m for $x>1$ m. The velocity is zero everywhere. The solution consists of a rarefaction, a stationary contact wave and a shock. Fig.~\ref{fig:SS_soluzione} shows the reference solution of the problem, in terms of free-surface elevation, computed according to \cite{LeFloch2007,LeFloch2011}.

\begin{figure}
\centering
\includegraphics[width=10cm]{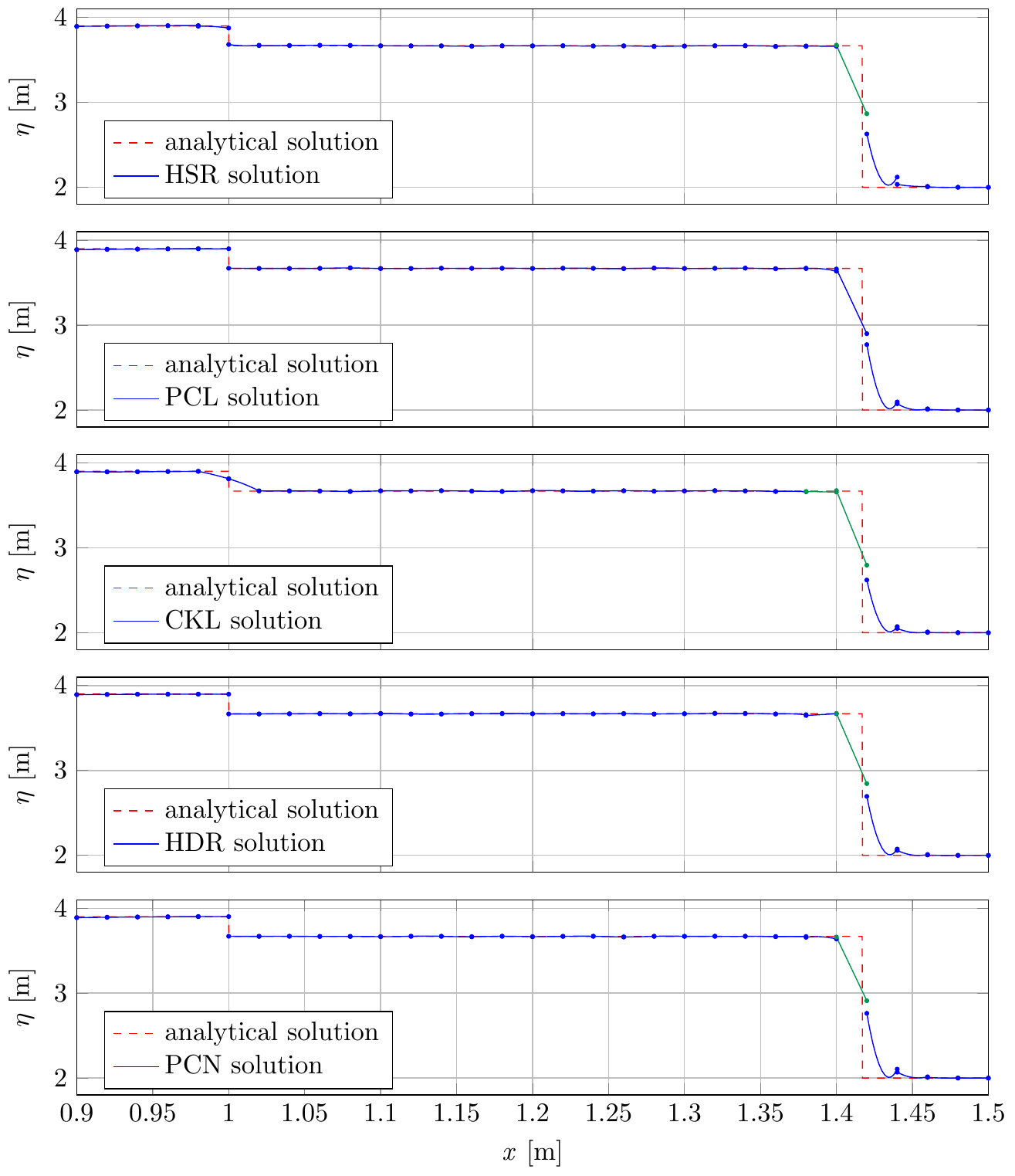}
\caption{Riemann problem with a bottom step: comparison between numerical and reference solutions for the water level. All the simulations are performed using 100 cells. Only the computational domain between 0.9 m and 1.5 m is shown. Green line are the free-surface level in the cells where the limiter is applied.}
\label{fig:SS_zoom}
\end{figure}

Fig.~\ref{fig:SS_zoom} shows the comparison between numerical and reference solutions for the water level. All the five models work well for this test case. Only the CKL model introduces an unphysical smooth transition between the water levels before and after the bottom step. This behavior is due to the restoration of the bottom continuity at the cell-interfaces that characterize the CKL approach. The good shock-resolution of the two path-conservative schemes, and in particular of the model with the non-linear path is an important achievement of the present work. In fact, it is well-known that the path-conservative models may poorly reproduce the shocks if the amplitude of such shocks are large \cite{Castro2008,Abgrall2010}. In particular, the model based on the new non-linear path shows shock-resolution properties as good as the classical model based on the linear path.

%

\subsection{Steady flow over a bottom step}
\begin{figure}[h]
\centering
\includegraphics[width=10cm]{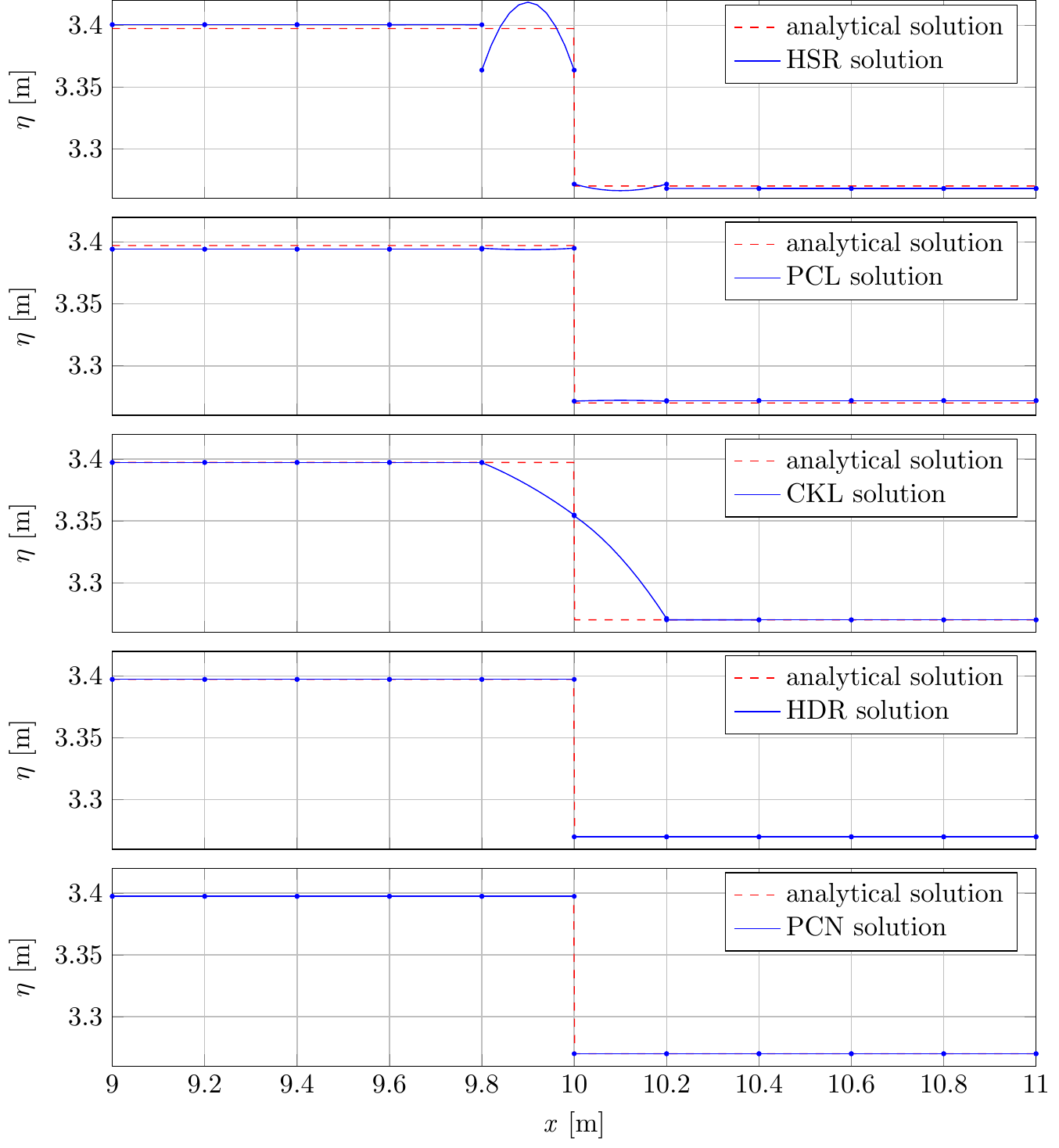}
\caption{Steady flow over a bottom step: comparison between numerical and analytical solutions for the water level. All the simulations are performed using 50 cells. Only the computational domain between 9 m and 11 m is shown. The limiter is not applied in the cells represented in the figure.}
\label{fig:LLL}
\end{figure}
This very simple test case is selected to verify the behavior of the models in simulating a steady flow over a bottom discontinuity. A flat channel with a single step, 1 m high, located at $x$ = 10 m, is considered. The computational domain is 20 m long. The flow is characterized by a total head $H$ equal to 3.5 m and a specific critical energy $E_{cr}$ equal to 2 m. The upstream discharge $q_*$ and the downstream sub-critical water depth $h_*$, used to impose the boundary conditions, are obtained satisfying the following relationships:
\begin{equation}
h_{cr} = \frac{2}{3}E_{cr}; \qquad h_{cr} = \sqrt[3]{\frac{q_*^2}{g}}; \qquad H = z + h_* + \frac{q_*^2}{2\,g\,h_*^2}.
\end{equation}
The initial, piecewise constant, moving water steady flow has to be preserved.

Fig.~\ref{fig:LLL} shows the comparison between the numerical solutions and the analytical free-surface elevation. Only the portion of the channel between $x=9$ m and $x=11$ m is represented in the figure. The HSR and PCL models are not able to correctly reproduce the steady jump in the water level induced by the step while the HDR, PCN and CKL models show a physically correct behavior. Moreover, it is also worth noting that the CKL model introduce an artificial smooth transition between the water levels before and after the step.

The classical hydrostatic reconstruction approach \cite{HSR}, applied in the model HSR, doesn't allow the correct reproduction of the water level discontinuity at the step. This fact doesn't surprise because the method is based on the correction of the flux related only on the static force exerted by the step, completely omitting the effect of the dynamic forces. These dynamical effects can be highlighted by a simple momentum balance over a control volume that includes the step. On the contrary, the model HDR \cite{Caleffi2009} is able to correctly reproduce the discontinuity because the flux correction takes into account the dynamical force exerted by the step on the flow and preserves the total head.

\subsection{Surge Crossing a Step}
\begin{figure}
\centering
\includegraphics[width=10cm]{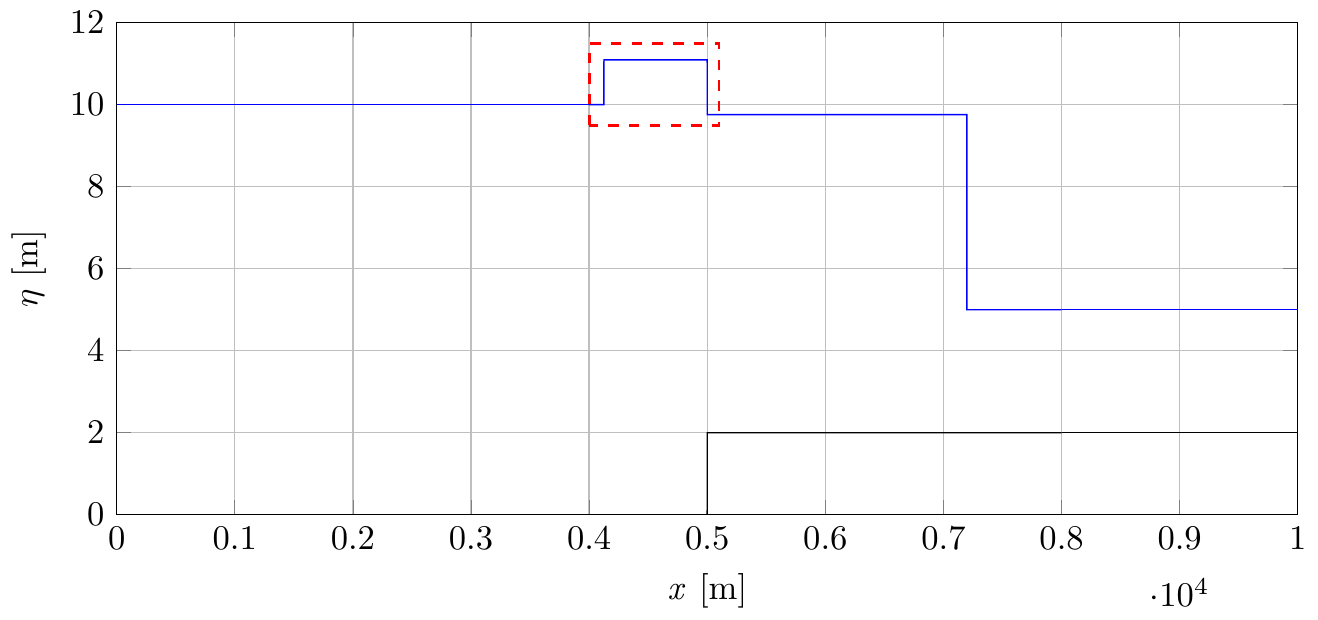}
\caption{Surge Crossing a Step: the solution is constituted by a shock, a stationary contact wave and a shock. Only the results inside the red rectangle are shown in Fig.~\ref{fig:USS}.}
\label{fig:UU_init}
\end{figure}
\begin{figure}
\centering
\includegraphics[width=10cm]{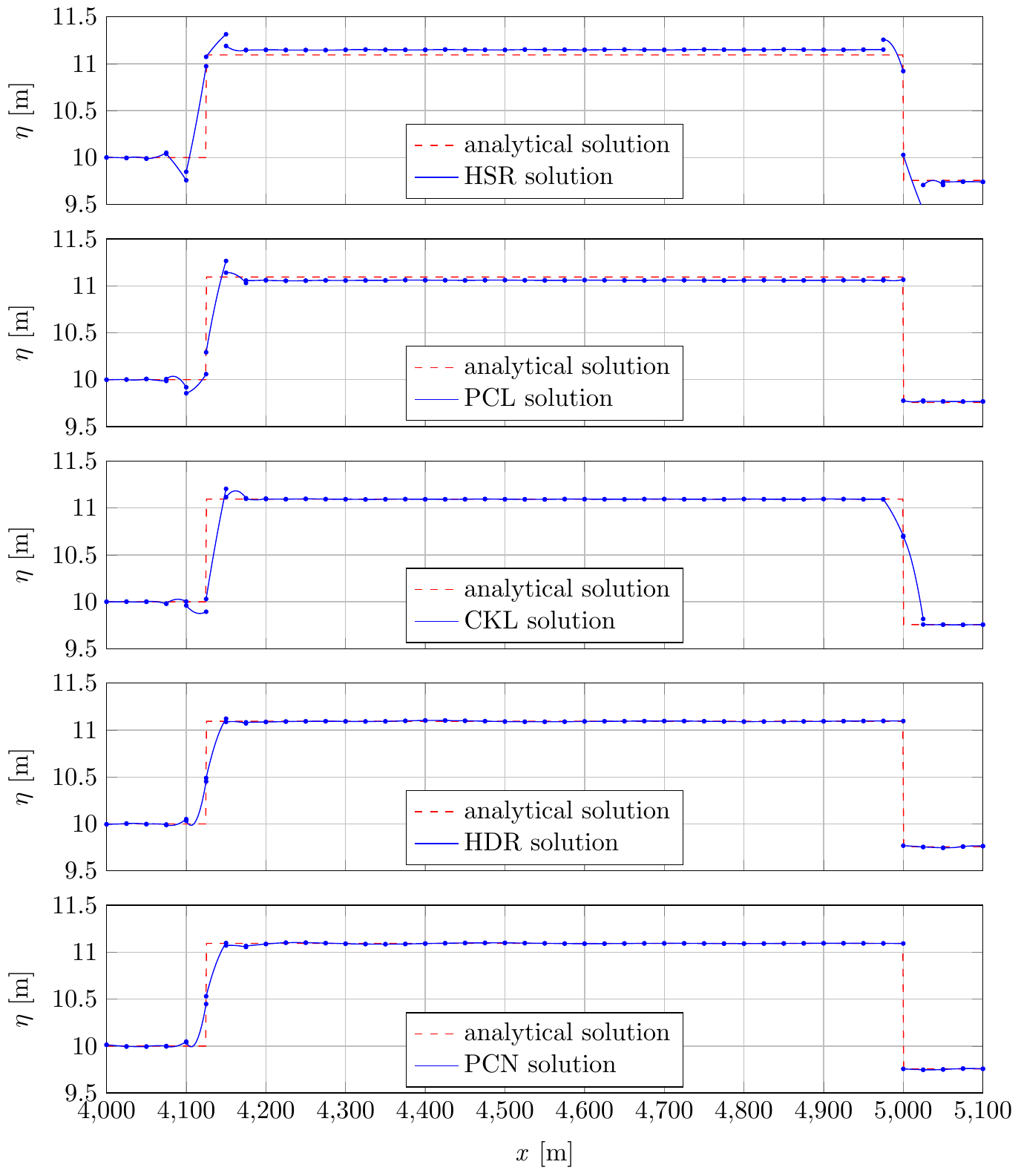}
\caption{Surge Crossing a Step: comparison between numerical and analytical solutions for the water level. All the simulations are performed using 400 cells. Only the computational domain between 4000 m and 5100 m is shown. The limiter is not applied in the cells represented in the figure.}
\label{fig:USS}
\end{figure}
This test case, conceived by Hu et al. \cite{hu2000nsw}, is used to verify the behavior of numerical models in the simulation of unsteady flow over discontinuous bottom. The channel is 10\,000 m long and the bottom elevation is zero for $x<5\,000$ m and 2 m for $x>5\,000$ m. The initial free-surface level is 5 m and the velocity is zero everywhere. The upstream boundary condition is characterized by a water depth of 10 m and by a flow velocity of:
\begin{equation}
v(0,t) = (\eta_u - \eta_d)\sqrt{\frac{g(\eta_u + \eta_d)}{2\eta_u \eta_d}};
\end{equation}
with $\eta_u$ = 10 m and $\eta_d$ = 5 m. The simulation time is $t$ = 600.5 s. The boundary conditions induces a surge that propagates downstream. When the surge reaches the bottom step, two surges are created, one moving upstream and one downstream. Fig.~\ref{fig:UU_init} shows the analytical solution. 

This unsteady flow is simulated using all the five models and the comparison of the obtained results are performed in terms of water elevation, similar results are obtained in terms of water discharge. 
Fig.~\ref{fig:USS} shows the solutions for the space interval $x \in [4\,000,\ 5\,100]$ m. The classical hydrostatic reconstruction approach \cite{HSR}, applied in the model HSR, doesn't allow the correct reproduction of the water level discontinuity at the step. Again this behavior can be explained remembering that the flux correction is related only on the static force exerted by the step on the flow and not to the dynamic forces. On the contrary, the model HDR \cite{Caleffi2009} is able to correctly reproduce the discontinuity because of the improved flux correction.
A similar reasoning can be applied to the couple of models based on a path-conservative approach (PCL and PCN models). While the use of a simple linear path doesn't allow the proper reproduction of the jump the non-linear path gives very satisfactory results. 
The simplest model CKL gives the right values of the jump strength across the step (located at $x$ = 5\,000 m) but the artificial reconstruction of the bottom continuity, obtained modifying the bottom slope of the cells near the step, leads to wrong values of the water elevation in the two cells with modified bottom slopes.

\section{Conclusions}
While the solutions for the well-balancing of a SWE model in the case of a quiescent flow are very numerous, few approaches for the well-balancing of a moving steady state are present in the literature. In this work we give a contribution to the well-balancing of SWE models for steady flow, indicating how some key elements of the standard approaches have to be changed to improve the overall behavior of the schemes. In particular, we have focused our attention on the treatment of the bottom discontinuity, both in the framework of the classical finite volume approach (suggesting the use of the hydrodynamic reconstruction instead of the hydrostatic reconstruction) and of the path-conservative schemes (suggesting the use of a specific curvilinear path in the computation of the fluctuations). Both these techniques are promising as proved by the results shown here. However, a further effort is needed to make these techniques applicable to a wider practical context. 

\end{document}